\documentclass[journal=nalefd,manuscript=letter]{achemso}
\usepackage{graphicx}
\usepackage{ragged2e}
\usepackage{mathtools}
\usepackage{physics}
\usepackage{xcolor}
\title{Fermi Velocity Dependent Critical Current in Ballistic  Bilayer Graphene Josephson Junctions}
\author{Amis Sharma}
\affiliation{Department of Physics and Astronomy, Texas A$\&$M University, College Station, 77843, Texas, USA}
\author{Chun-Chia Chen}
\affiliation{Department of Physics, Duke University, Durham, 27701, North Carolina, USA}
\author{Jordan McCourt}
\affiliation{Department of Physics, Duke University, Durham, 27701, North Carolina, USA}
\author{Mingi Kim}
\affiliation{Department of Physics and Astronomy, Purdue University, West Lafayette, 47907, Indiana, USA}
\author{Kenji Watanabe}
\affiliation{Advanced Materials Laboratory, NIMS, Tsukuba, 305-0044, Japan}
\author{Takashi Taniguchi}
\affiliation{Advanced Materials Laboratory, NIMS, Tsukuba, 305-0044, Japan}
\author{Leonid Rokhinson}
\affiliation{Department of Physics and Astronomy, Purdue University, West Lafayette, 47907, Indiana, USA}
\author{Gleb Finkelstein}
\affiliation{Department of Physics, Duke University, Durham, 27701, North Carolina, USA}
\author{Ivan Borzenets}
\affiliation{Department of Physics and Astronomy, Texas A$\&$M University, College Station, 77843, Texas, USA}
 \email{borzenets@tamu.edu}

\begin{document}
\begin{abstract}
    We perform transport measurements on proximitized, ballistic, bilayer graphene Josephson junctions (BGJJs) in the intermediate-to-long junction regime ($L>\xi$). We measure the device's differential resistance as a function of bias current and gate voltage for a range of different temperatures.  The extracted critical current $I_{C}$ follows an exponential trend with temperature: $ \exp(-k_{B} T/ \delta E)$. Here $\delta E = \hbar \nu_F /2\pi L $: an expected trend for intermediate-to-long junctions. From $\delta E$, we determine the Fermi velocity of the bilayer graphene, which is found to increase with gate voltage. Simultaneously, we show the carrier density dependence of $\delta E$, which is attributed to the quadratic dispersion of bilayer graphene. This is in contrast to single layer graphene Josephson junctions, where $\delta E$ and the Fermi velocity are independent of the carrier density.  The carrier density dependence in BGJJs allows for additional tuning parameters in graphene-based Josephson Junction devices.
\end{abstract}
\maketitle

Ballistic graphene Josephson junctions (GJJs) have been widely utilized as a platform to study novel quantum physics phenomena\cite{Park2024, Borzenets2016-2} and devices\cite{Kroll2018}, including: entangled pair generation\cite{Chen2015, Borzenets2016}, topological states arising from the mixing of superconductivity and quantum Hall states\cite{Amet2016}, as well as photon sensing via bolometry/calorimetry\cite{Lee2020}. Superconductor-normal metal-superconductor Josephson junction (SNSJJ) hosts Andreev bound states (ABS), which carry supercurrents across the normal region of the JJ; in order to enter the ballistic regime, a disorder-free weak link and high transparency at the SN interface are necessary. Hexagonal Boron-Nitride (hBN) encapsulated graphene as a weak link enables highly transparent contacts at the interface whilst keeping graphene clean throughout the fabrication process \cite{Dean2010}. Here, we study proximitized, ballistic, bilayer graphene Josephson junctions (BGJJs). Bilayer graphene devices (in contrast to monolayer) allow extra potential tunability via a non-linear dispersion relation, applied displacement field, or lattice rotation \cite{Park2024}.

The critical current ($I_{C}$) of SNSJJ in the intermediate-to-long regime, where the junction length (L) $\geq$ superconducting coherence length ($\xi_{0}$), scales with temperature (T) as $I_{C}= exp(-k_{B} T/ \delta E)$. Here, $\delta E = \hbar \nu_{F}/2\pi L$, an energy scale related to the ABS level spacing \cite{Borzenets2016-2,Kulik1970,Bardeen1972,Svidzinsky, Svidzinsky-2, Bagwell}. Note that in the intermediate regime ($L\approx\xi_0$)  $\delta E$ is found to be suppressed \cite{Borzenets2016}. A previous study of GJJs found that in this regime, the relation is held more precisely when $\xi$ was taken into account along with L, that is: $\delta E = \hbar \nu_{F}/2\pi (L+\xi)$ \cite{Borzenets2016-2, Bagwell}.  Monolayer graphene displays a linear dispersion relation, which results in a constant fermi velocity ($\nu_{F0}$). Thus, in ballistic GJJs, $\delta E$ remains independent of the carrier density. In comparison, bilayer graphene displays a quadratic dispersion relation at low energies.  In BGJJs we studied, a back-gate voltage ($V_G$) controls the carrier density; and $\delta E$ dependence on $V_G$ is observed. Using $\delta E$, we extract the Fermi velocity in bilayer graphene: It is seen that $\nu_F$ increases with $V_G$, and saturates to the constant value, $\nu_{F0}$, of the monolayer graphene.

Our device consists of a series of four terminal Josephson junctions (on $\mathrm SiO_2/Si$ substrate) made with hBN encapsulated bilayer graphene contacted by Molybdenum-Rhenium (MoRe) electrodes. Bilayer graphene is obtained via the standard exfoliation method. It is then encapsulated in hexagonal Boron-Nitride using the dry transfer method \cite{Wang2013}.  MoRe of $80$ nm thickness is deposited via DC magnetron sputtering. The resulting device has four junctions of lengths $400$ nm, $500$ nm, $600$ nm, and $700$ nm. The width of the junctions is  $4~\mu$m. The device is cooled in a Leiden cryogenics dilution refrigerator operated at temperatures above $1$ K, and measurements were performed using the standard four-probe lock-in method. A gate voltage $V_G$ is applied to the $Si$ substrate with the oxide layer acting as a dielectric, which allows modulation of the carrier density. \cite{Borzenets2011,Borzenets2013,Borzenets2016,Borzenets2016-2,Tang2022,Amet2016}.
\begin{figure}[h]
    \centering
    \includegraphics[width=0.9\linewidth]{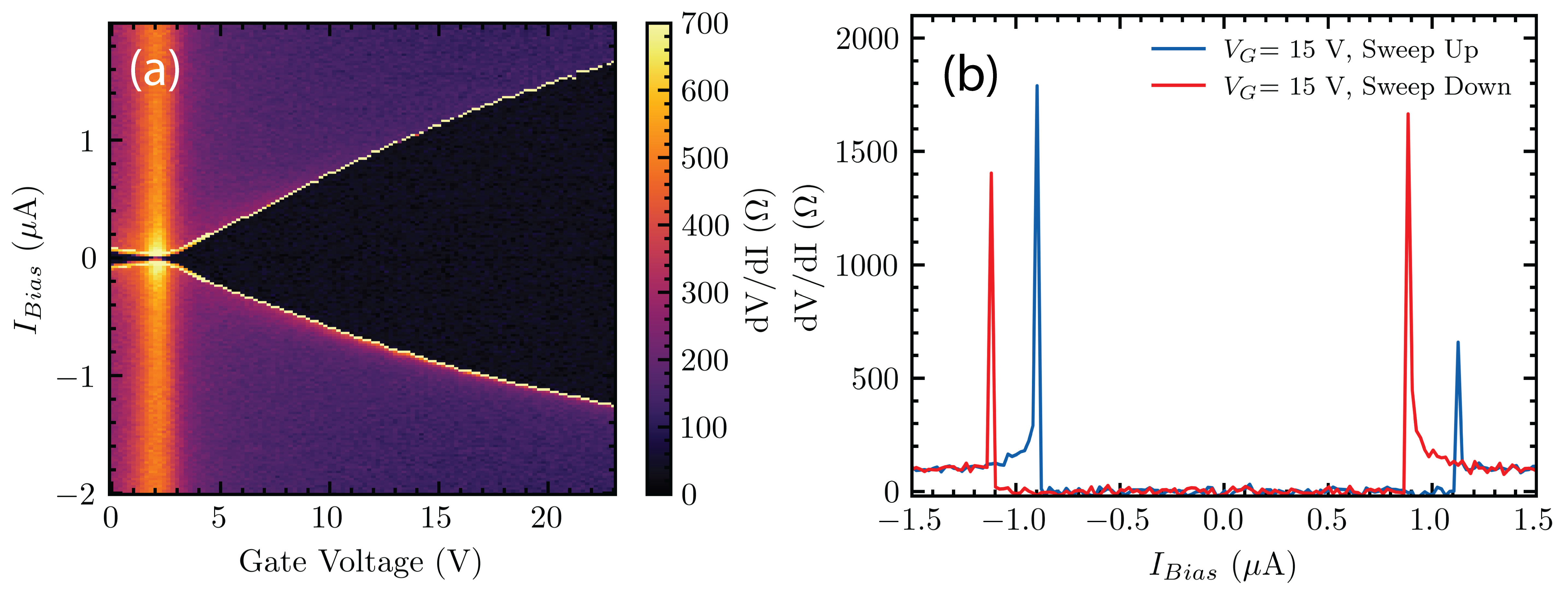}
    \caption{(a) Differential resistance ($dV/dI$) versus gate voltage ($V_G$) and bias current $I_{Bias}$ taken at $T=1.37$ K. The black region around zero bias corresponds to the superconducting state. $I_{Bias}$ is swept up (from negative to positive). Thus, the transition at negative bias corresponds to the re-trapping current $I_R$, while the transition at positive bias is the switching current $I_C$. (b) Vertical line cut of the resistance map taken at $V_G=15$ V, $T=1.37$ K,  showing device's $dV/dI$ versus bias current. Blue line corresponds to $I_{Bias}$ swept up, with red line swept down (positive to negative).  }
\end{figure}
Figure. 1(a) displays the differential resistance ($dV/dI$) map of the $400$ nm junction at $T=1.37$ K; we see zero resistance (black region) across all applied $V_G$ indicating the presence of supercurrent. As the bias current $I_{bias}$ is swept from negative to positive values, the junction first reaches its superconducting state at a value $|I_{bias}|= I_R$, known as the re-trapping current. Then, as $|I_{bias}|$ is increased to higher positive values, the junction transitions to the normal state at $|I_{bias}|= I_S$, known as the switching current. Figure. 1(a) shows that the junction can sustain a larger region of critical current as we modulate the carrier density to higher values via $V_G$. Fig. 1(b) displays line traces extracted from the $dV/dI$ map which shows hysteresis in $I_R$ and $I_S$. This is a commonly observed phenomenon in underdamped junctions \cite{tinkham,Borzenets2011},  or can also be attributed to self-heating \cite{Borzenets2013,Courtois2008,Tang2022}. The measured switching current $I_S$ is slightly suppressed compared to the junction's ``true" critical current $I_C$. However, previous measurements on the statistical distribution of $I_S$ in similar graphene devices found that $I_S$ is suppressed from $I_C$ by no more than $10\%$ for critical currents up to a few $\mu A$ \cite{Borzenets2016-2, Coskun2012,Lee2011,Ke2016}.

Extracting the critical current $I_C$ from the differential maps for different temperatures, we can see that $I_C$ falls exponentially with inverse $T$  (Figure. 2c) We also extract the conductance of the junction in the normal regime ($I_{Bias}\gg I_C$). Figure. 2(b) shows this conductance ($G$) for the $400$ nm junction device. Due to the significant contact resistance ($R_C$) of the device, the measured conductance $G$ is uniformly suppressed compared to the ballistic limit expectation. However, when accounting for $R_C$ within the fit, we find that the conductance $G$ scales as the square-root (as opposed to linearly) of $V_G$ (blue curve of Figure 2(b)). This is consistent with ballistic transport\cite{Graphene_Review,Borzenets2016-2}. To further demonstrate the ballistic nature of the device, we present normal resistances ($R_N$) of junctions of length $500$ nm, $600$ nm, and $700$ nm with the fitted, constant contact resistance $R_C$ subtracted (Figure. 2(b) inset).  The inset plot shows that the values of $R_N - R_C$ are independent of the junction length, demonstrating the ballistic nature of the devices.
\begin{figure}[ht]
    \centering
    \includegraphics[width=0.9\linewidth]{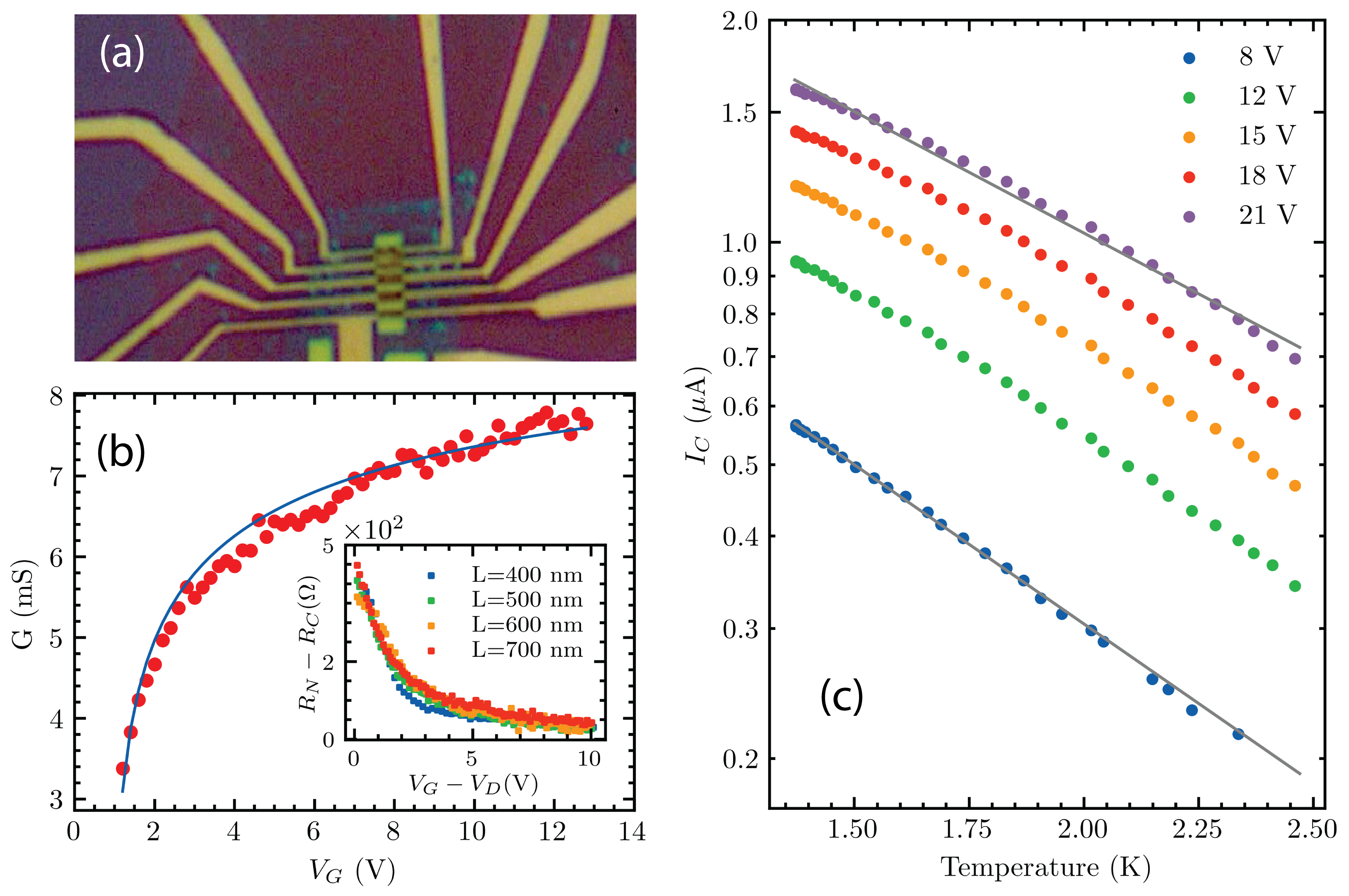}
    \caption{(a) Device picture. Image shows series of junctions with different Lengths: $400$ nm, $500$ nm, $600$ nm and $700$ nm. (b) The ballistic conductance vs Gate voltage for $L=400$ nm junction. The blue curve corresponds to the fit for ballistic devices, with an addition of a contact resistance. The inset shows junction resistance minus the parasitic contact resistance plotted against gate voltage from the Dirac point for all our devices. (c) Critical currents $I_C$ of $L=400$ nm junction plotted against temperature $T$, for various gate voltages, on a semi-log scale. The plots show $V_G$ dependence of $I_C$: the gray lines show that the slope of the curve for the lowest plotted gate $V_G=8$ V, is smaller than the slope of the highest plotted gate $V_G=21$ V.  } 
\end{figure} 

To extract $\delta E$ of the junction, we go to the discussion of $I_C$ vs. the temperature trends in Figure. 2(c).  Here, the y-axis is plotted in logarithmic scale. From the slope of the curves $\mathrm{Log}(I_C) = - (k_B/ \delta E) T$ for each gate, one can extract $\delta E$ versus $V_G$ (plotted in Figure. 3a). Unlike for the case of monolayer graphene, a clear dependence on $V_G$ is seen (The observed trend further supports the view that our devices operate in the long ballistic regime. Diffusive Josephson junctions are governed by the Thouless energy $E_{Th} \propto 1/[(R_N-R_C) \sqrt {V_G-V_D}]$ \cite{Ke2016,Diffusive} which does not match the trend with respect to $V_G$ seen in Figure. 3(a) ). The energy $\delta E$ scales linearly with the Fermi velocity $v_F$ (Figure. 3(b) ). Note that calculating $v_F$ from $\delta E$ for junctions in the intermediate regime requires knowledge of the superconducting coherence length $\xi$. In the fit discussed below, we use $\xi$'s dependence in $v_F$.

We now compare the experimentally obtained $\delta E$ (and $v_F$) to the theoretical expectation.  With the dispersion relation for bilayer graphene written as : $\mathcal{E} = \frac{1}{2} \gamma_{1} ( \sqrt{ 1+ 2 \hbar^2 k^2/ \gamma_{1}^2 m^*} -1) $, we get the expression for the Fermi velocity : $v_{F}= \sqrt{\frac{2\mathcal{E}_{F} \gamma_1 (\mathcal{E}_{F}+\gamma_1)}{(2\mathcal{E}_{F}+\gamma_1)^{2}m^*}} $ \cite{McCann2013,Fang2007,Fates2019}. Here,  $\gamma_1=0.39~eV$ a parameter describing the interlayer coupling \cite{McCann2013}, $k$ is the momentum wavevector, $m^*$ is the effective mass of electrons. Moreover, the Fermi energy $\mathcal{E}_{F}$ for bilayer graphene scales as: $\mathcal{E}_{F} = \frac{\hbar^{2} \pi |n|}{2 m^{*}}$.  The carrier concentration $n$,  controlled by the applied gate voltage $V_G$, is given by $n = \frac{V_{G} - V_{D}}{e} C_{Total} $ with $V_D$ as the gate voltage at the Dirac point. The total capacitance $C_{Total}$ is a combination of quantum capacitance $C_q$ and gate oxide capacitance $C_{ox}$:  $C_{Total} =\left[ \frac{1}{C_{ox}}+\frac{1}{C{q}} \right]^{-1} $. The quantum capacitance $C_q$ for bilayer graphene is determined by $C_{q}=\frac{ 2 e^{2} m^{*}} {\pi \hbar^{2}}$, where $e$ is the electron charge. The gate oxide capacitance per unit area is $ C_{ox} = \frac{\epsilon_0 \epsilon_{r}}{d}$, where $\epsilon_{0}$ is the vacuum permittivity, $\epsilon_{r}$ is the relative permittivity of the oxide, and $d$ is the thickness of the oxide layer. For silicon oxide gate with $d=300$ nm we get $C_{ox}\approx 115 \mu F/m^2$. Thus, the full expression for the Fermi velocity $v_F$ is: \\
\begin{align}
v_{F}= \hbar \sqrt{\frac{2\pi e \epsilon_0 \epsilon_r\gamma_1 (V_G-V_D)(2 d e^2\gamma_1 m^* + \pi \epsilon_0\epsilon_r\hbar^2(e(V_G-V_D)+\gamma_1))} {m^*(2 d e^2 \gamma_1 m^* + \pi \epsilon_0 \epsilon_r \hbar^2(2 e(V_G-V_D) + \gamma_1))^2 }}
\end{align}
Note that the effective mass $m^*$ typically ranges from $0.024~m_e$ to $0.058~m_e$ for $1*10^{12} \sim 4*10^{12}$ carriers/$cm^2$ \cite{Zou2011}, where $m_e$ is the electron rest mass.
Experimental data provides us with the following: $\delta E(V_G)= \frac{\hbar}{2\pi(L+\xi)}v_F$. We also note that $\xi$ has a dependence on $v_F$ and the superconducting gap $\Delta$: $\xi=\hbar v_F /2\Delta$\cite{Bagwell}. To fit $\delta E$, the model is set as: $\delta E(V_G)=\mathcal{F}(m^*, \Delta, V_D, d )$ where $m^*, \Delta, V_D, d$ are the fitting parameters, and $V_G$ is the independent variable. (We use the as-designed length of the device $L$, and take $\epsilon_{r}=3.9$ for $SiO_{2}$.)
\begin{table}[ht]
    \centering
\begin{tabular}{ |p{3cm}|p{3cm}|p{4cm}|}
      \hline
      Parameter & Fitted Value & Expected Value\\
      \hline
      $\Delta$ & $0.99$ meV & $0.8\sim1.2$ meV\\ 
      \hline 
      $d$ & $323$ nm  & $280\sim330$ nm\\
      \hline
      $m^*$ & $0.028~m_e$ & $0.02\sim0.06~m_e$\\
      \hline
      $V_D$ & $2.04$ V & $\approx +2$ V\\
      \hline
\end{tabular}
\caption{The fitting parameters used to match the measured $\delta E$, and consequently the Fermi velocity $v_F$, versus gate to the theoretical expectation described in Equation 1.  We see that resulting fitted values match closely to what is expected. The expected gate dielectric thickness $d$ is estimated from the substrate specifications plus the bottom hBN thickness. The expected Dirac point voltage $V_D$ is obtained from the resistance map. The expectations for superconducting gap $\Delta$ and the effective mass $m_i$ are obtained from previous works\cite{Zou2011, Borzenets2016}.}
\end{table}

The resulting fits of the data from the $400~nm$ junction for $\delta E$ and $v_F$ are plotted as solid lines in Figure. 3(a) and Figure. 3(b) respectively. Moreover, taking the fitted parameters from Table 1, we calculate the Fermi velocity $v_F$ for the available data points of all other junctions on the same substrate. As seen from Figure. 3(b), the calculated $v_F$ of all devices is in good agreement with the fit obtained from the $400$ nm junction (This is as expected for devices on the same substrate; as long as they have consistent parasitic doping and superconductor-graphene contact interface). The fitted parameters are summarized in Table 1. All fall within the range of expected values, with $\Delta$ being consistent with previously measured values for graphene/MoRe junctions\cite{Borzenets2016-2}. Furthermore, using the values obtained from the model, we find that $v_F$ saturates to the value of $1.1*10^6$ m/s as $V_G$ tends to infinity. \\
\begin{figure}[ht!]
     \centering
     \includegraphics[width=0.48\linewidth]{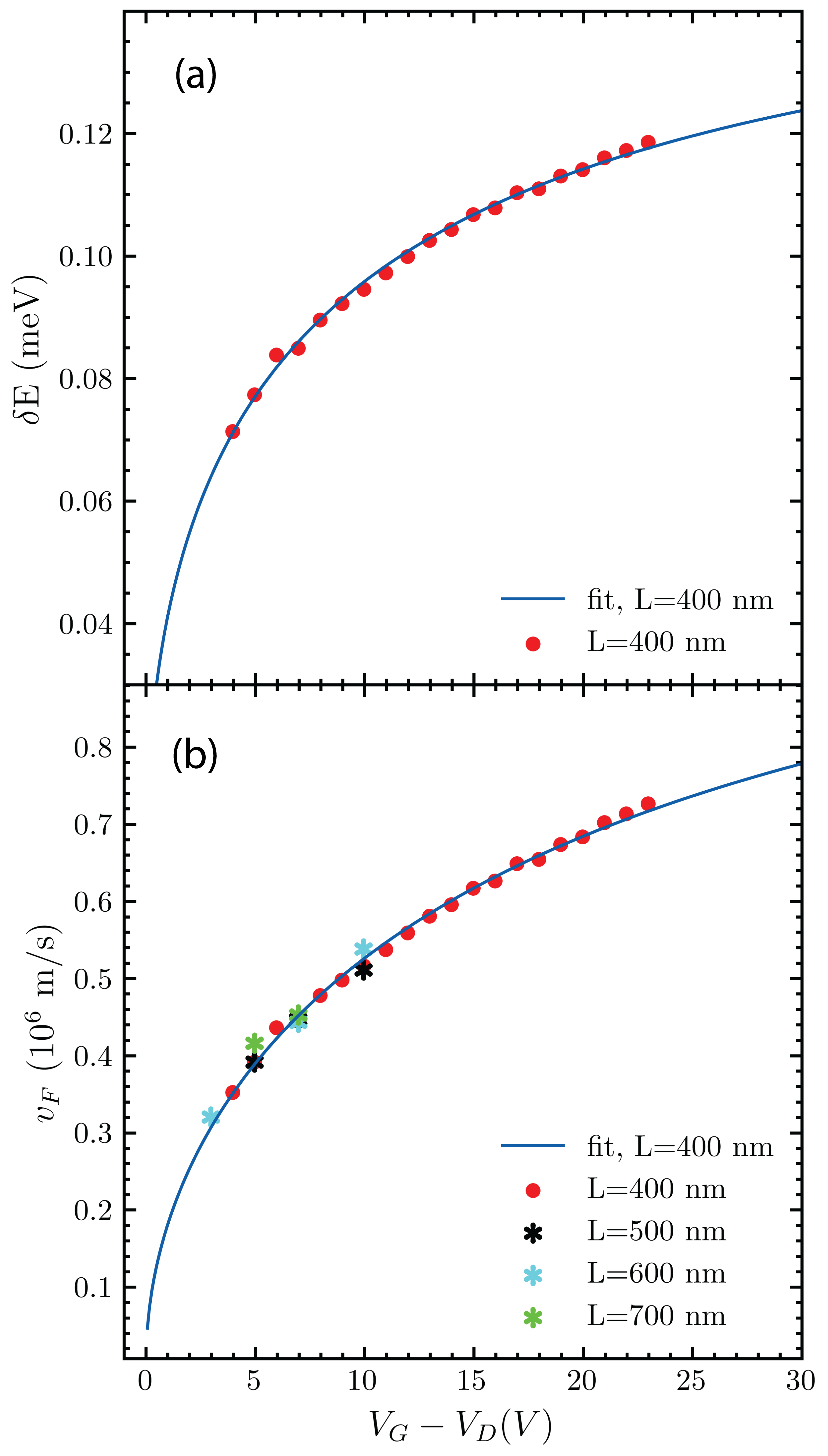}
     \caption{ (a) Energy $\delta E$ extracted from the slope of log($I_C$) vs T plotted against the gate voltage $V_G$ from the Dirac point of the junction with $L=400$ nm. We see $\delta E$ dependence on the carrier density modulated via the gate voltage for the junction. (b) Fermi velocity ($v_f$) calculated from $\delta E$ using the device dimensions, and parameters obtained from the fit to theory. The solid line represents the theoretical trend as fitted to the data for the $L=400$ nm junction. In addition, panel (b) shows calculated $v_F$ for the other junctions using parameters obtained from the $L=400$ nm fit.}
 \end{figure}
In conclusion, we study the evolution of the critical current with respect to the gate in bilayer graphene Josephson Junctions (BGJJs). Using the critical current-temperature relation expected for intermediate-to-long junctions, we extract the relevant energy scale $\delta E$ and find that it has a clear gate dependence. As $\delta E$ is proportional to the Fermi velocity $v_F$ in bilayer graphene, we are able to match the observed gate dependence to the theoretical expectation. Our observation is contrasted with monolayer graphene JJs, which do not have a gate-dependent $\delta E$. This result showcases the greater tunability of BGJJs, and offers additional avenues for device characterization. Although not observed here, it should be possible to engineer Josephson junctions that transition from the short to the intermediate/long ballistic regimes in-situ via gate voltage. The ability to tune ABS level spacing could have applications in self-calibrating sensors, or for matching resonance conditions in multi-terminal superconducting devices.

L.R. acknowledges support from NSF (DMR-2005092 award) for contact deposition. Lithographic fabrication and characterization of the samples were performed by J.M. and supported by the Division of Materials Sciences and Engineering, Office of Basic Energy Sciences, U.S. Department of Energy, under Award No. DE-SC0002765. Measurements conducted by C.-C. C., and data analysis performed by G.F. were supported by the NSF Award DMR-2428579. I.V.B. and A.S. acknowledge the support from Texas A$\&$M University.
\clearpage
\bibliography{main}
\end{document}